\documentclass[prb,twocolumn,aps,showpacs]{revtex4}
\usepackage{amsfonts}
\usepackage{graphicx}
\usepackage{textcomp}
\usepackage{txfonts}
\begin{document}
\title{First Principles Studies on 3-Dimentional Strong Topological
  Insulators: Bi$_2$Te$_3$, Bi$_2$Se$_3$ and Sb$_2$Te$_3$}

\author{Wei Zhang, Rui Yu, Hai-Jun Zhang, Xi Dai and Zhong Fang}

\affiliation{Beijing National Laboratory for Condensed Matter
Physics, and Institute of Physics, Chinese Academy of Sciences,
Beijing 100190, China;}

\date{\today}   
\begin{abstract}

Bi$_2$Se$_3$, Bi$_2$Te$_3$ and Sb$_2$Te$_3$ compounds are recently
predicted to be 3-dimentional (3D) strong topological insulators. In
this paper, based on {\it ab-initio} calculations, we study in detail
the topological nature and the surface states of this family
compounds. The penetration depth and the spin-resolved Fermi surfaces
of the surface states will be analyzed.  We will also present an
procedure, from which highly accurate effective Hamiltonian can be
constructed, based on projected atomic Wannier functions (which keep
the symmetries of the systems). Such Hamiltonian can be used to study
the semi-infinite systems or slab type supercells
efficiently. Finally, we discuss the 3D topological phase transition
in Sb$_2$(Te$_{1-x}$Se$_x$)$_3$ alloy system.

\end{abstract}


\maketitle

\section{ Introduction }

The topological insulators
(TIs)~\cite{Kane_PRL95_2005_146802,Bernevig_PRL96_2006,Moore,Fu1,Fu2},
are interesting not only because of its fundamental importance but
also because of its great potential for future
applications~\cite{Majorana,MEC,MEC_vanderbilt}. TI is a new state of
quantum matter, and is distinct from simple metal or insulator in the
sense that its bulk is insulating (with a bulk band gap), while its
surface (or edge) is metallic due to the presence of gapless surface
(edge) states. Those surface states are spin splitted but with double
degeneracy at the Dirac point, which is protected by the time-reversal
symmetry. The surface states, consisting of odd number of massless
Dirac cones, are robust against time-reversal-invariant perturbations
and also are very different with Graphene where spin degeneracy is
reserved (the concept of pseudo-spin describing the sub-lattice
degeneracy is used for Graphene). At the early stages of studies for
TIs, due to the absence of realistic materials, most of the
discussions were based on the model
Hamiltonians~\cite{Kane_PRL95_2005_146802,Bernevig_PRL96_2006,Moore,Fu1,Fu2}.
However, within last couple of years, this field is strongly motivated
by the discoveries of several real compounds.

Two-dimension (2D) TI (or called quantum spin Hall insulator) has been
proposed and realized in HgTe/CdTe quantum well structure through
varying the well thickness~\cite{Bernevig,Konig,Dai}. Three-dimension
(3D) topological nature of Bi$_{1-x}$Sb$_x$ alloy has been
demonstrated by fine tuning of the alloy
concentration~\cite{Hsieh,Kane78_2008,Zhang2}. In both cases, the bulk
band gap is of the order of tens meV, too small for the potential
applications. In our recent studies, however, we predicted that
Bi$_2$Se$_3$ family compounds~\cite{Zhang} (i.e., Bi$_2$Te$_3$,
Bi$_2$Se$_3$ and Sb$_2$Te$_3$), which are well-known thermoelectric
materials~\cite{Thermo,Mishra,Larson}, are strong topological
insulators with surface states consisting of single Dirac cone at the
$\Gamma$ point. Particularly, for Bi$_2$Se$_3$, a bulk band gap of 0.3
eV is predicted, much larger than the energy scale of room
temperature.  The compounds are stoichiometric, chemically very
stable, and easy to be synthesized, and yet its surface states are
very simple. The existence of such surface states in this family
compounds have been experimentally confirmed on samples ranging from
bulk to thin film~\cite{Xia,chen,Kehui,Xue1,Xue2,Xue3}.

More and more studies are now going on for Bi$_2$Se$_3$ family
compounds, quantitative studies and comparison with experiments become
more and more important. On the other hand, fully self-consistent {\it
  ab~initio} calculations are quite limited due to the large system
size. In order to study both the bulk and the surface behavior
simultaneously, which is necessary to identify its topological nature,
a large slab type supercell is needed, and heavy calculations are
involved. To simplify future studies, it is therefore desirable to
have a highly accurate effective Hamiltonian, which will be one of the
main purposes of this paper.  It's well-known that the maximally
localized Wannier functions (MLWF) method introduced by Marzari and
his coworkers~\cite{Marzari,Souza} has played an important role in
constructing the effective Hamiltonian.  However, the algorithm
involved does not hold the symmetry when the MLWF are constructed from
Bloch functions. To solve this problem, we introduce a set of
projected atomic Wannier functions (PAWF), whose symmetry can be
easily controlled, yet keeping high accuracy. With this set of PAWF
basis, highly accurate effective Hamiltonian can be constructed. Using
this Hamiltonian, we will demonstrate that the topological nature as
well as the details of surface states can be all well reproduced.

The organization of the present paper is as follows. In section II, we
describe the crystal structure and symmetries of Bi$_2$Se$_3$ family
compounds. Section III is devoted to the construction of the effective
Hamiltonian in the basis of PAWF. In section IV, this Hamiltonian is
applied to Bi$_2$Se$_3$ system, and the topological nature and surface
properties are analyzed. In section V, we discuss the 3D topological
phase transition by doping Se into Sb$_2$Te$_3$. Final remarks are
summarized in section VI.

\section{Crystal structure and symmetries}

\begin{figure}[tbp]
\includegraphics[clip,width=7cm]{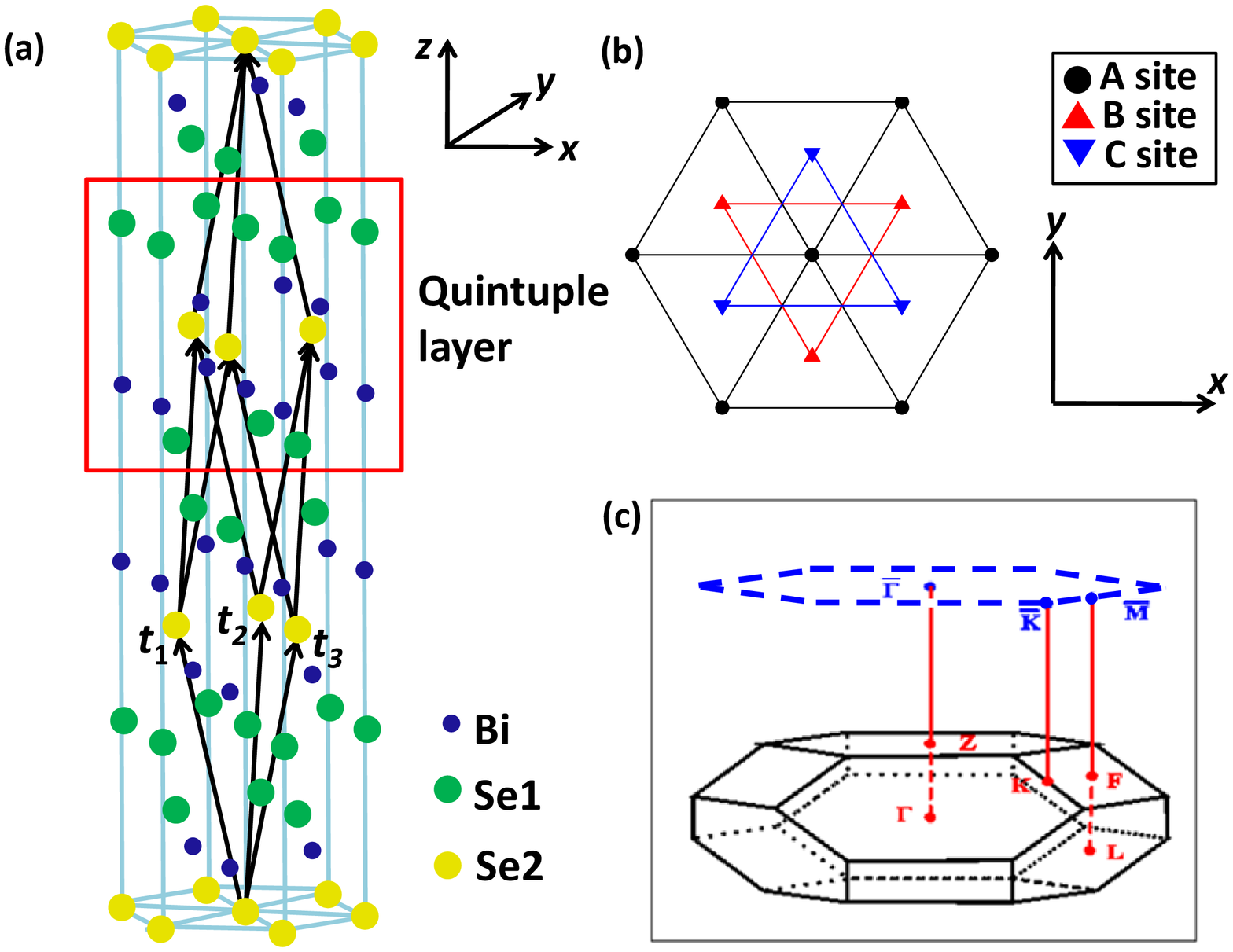}
\caption{(Color online) Crystal structure of Bi$_2$Se$_3$ family
  compounds.  (a) The hexagonal supercell containing 15 atomic layers
  and primitive translation vectors $\bf{t}_{1, 2, 3}$ (b) The top
  view of a quintuple layer (QL) in the triangle lattice. Three sets
  of different sites, labeled as A, B, C sublattices respectively, are
  presented. Owing to the $\bf{D_{3d}^{5}}$ symmetry, the stacking of
  atomic layers along the $z$ direction is in the order of
  $\cdots-$C(Se1)$-$A(Se1)$-$B(Bi1)$-$C(Se2)$-$A(Bi1)$-$B(Se1)$-$C(Se1)$\cdots$. (c)
  The first Brillouin zone (BZ). Four nonequivalent
  time-reversal-invariant-momentum (TRIM) points $\Gamma$ (0, 0, 0),
  $L (\pi, 0, 0)$, $F (\pi, \pi, 0)$, and $Z (\pi, \pi, \pi)$ are
  denoted in the 3D BZ.  The corresponding surface 2D BZ is
  represented by the dashed blue hexagon, and $\overline{\Gamma}$,
  $\overline{M}$, $\overline{K}$ are the corresponding TRIM special k
  points in the surface BZ.}\label{figBrillouin}
\end{figure}

The Bi$_2$Se$_3$ family compounds have the rhombohedral crystal
structure with space group $\bf{D_{3d}^{5}}$ ($\bf{R\bar{3}m}$), we
take Bi$_2$Se$_3$ as an example in the following. As shown in
Fig.~\ref{figBrillouin}(a), the system has a layered structure with
five atomic layers as a basic unit (cell), named a quintuple layer
(QL). The inter-layer bonding within the QLs is strong because of the
dominant covalent character, but the bonding between the QLs is much
weaker due to the van der Walls type interaction. The binary (with
twofold rotation symmetry), bisectrix (appearing in the reflection
plane), and trigonal (with threefold rotation symmetry) axes are taken
as x, y, z axes respectively, and the primitive translation vectors
$\bf{t}_{1, 2, 3}$ shown in Fig.~\ref{figBrillouin} are

\begin{eqnarray}
\nonumber
  \bf{t}_1 &=& (-a/2, -\sqrt{3}a/6, c/3),\\
  \bf{t}_2 &=& (a/2, -\sqrt{3}a/6, c/3),\\
  \nonumber
  \bf{t}_3 &=& (0, \sqrt{3}a/3, c/3).
\end{eqnarray}
Here, $a$ and $c$ are lattice constants in hexagonal cell.  The
corresponding reciprocal vectors $\bf{s}_{1, 2, 3}$, defined by
$\bf{s}_i\cdot\bf{t}_j=2\pi\delta_{ij}$, are given as,
\begin{eqnarray}
  \nonumber
\bf{s}_1 &=& (-1, -\sqrt{3}/3, b)h,\\
 \bf{s}_2 &=& (1, -\sqrt{3}/3, b)h,\\
  \nonumber
\bf{s}_3 &=& (0, 2\sqrt{3}/3, b)h,
\end{eqnarray} with
\begin{eqnarray}
\nonumber
b &=& a/c,\\
h &=& 2\pi/a.
\end{eqnarray}

As shown in Fig.~\ref{figBrillouin}(a), we take Se2 to be at origin
(0, 0, 0), then two Bi sites are at ($\pm\mu$, $\pm\mu$, $\pm\mu$),
and two Se1 are at ($\pm\nu$, $\pm\nu$, $\pm\nu$), defined in the unit
of primitive translation vectors. All the experimental lattice
parameters and internal parameters $\mu$ and $\nu$ are listed in
Table~\ref{parameters}.  Fig.~\ref{figBrillouin}(c) shows the 3D first
Brillouin zone (BZ) and the 2D surface BZ of Bi$_2$Se$_3$. $\Gamma(0,
0, 0)$, $L(\pi, 0, 0)$, $F(\pi, \pi, 0)$ and $Z(\pi, \pi, \pi)$ are
four time-reversal invariant momentum (TRIM) points in the 3D
BZ. $\Gamma(0,0,0)$ and $Z(\pi,\pi,\pi)$ are projected as
$\overline{{\Gamma}}$, and $L(\pi,0,0)$ and $F(\pi,\pi,0)$ are
projected as $\overline{M}$ in the surface BZ.  For the choice of our
cell, Bi$_2$Se$_3$ has the inversion symmetry with inversion center at
Se2.  The space group $\bf{R\bar{3}m}$ can be constructed from three
symmetry generators: I (inversion), $c_{3z}$ (three fold rotation
around $z$), and $\sigma_x$ (mirror plane with its normal along $x$).

\begin{table}
\caption{Experimental lattice parameters and the internal parameters
of Bi$_2$Se$_3$ family compounds~\cite{Wyckoff}.}
\begin{tabular}{l c c c c }
\hline
         & &Sb$_2$Te$_3$  &Bi$_2$Te$_3$   &Bi$_2$Se$_3$  \\

\hline

Lattice constant    &a (${\AA}$) &4.250  &4.383  &4.138     \\
                                      &c (${\AA}$) &30.35   &30.487 &28.64  \\
inner coordinates              &$\mu$     &0.400    &0.400 &0.399\\
                               &$\nu$      &0.211   &0.212 &0.206\\
 \hline
\end{tabular}\label{parameters}
\end{table}

\section{Effective Hamiltonian from PAWF}

The construction of the PAWF basis and the effective Hamiltonian is
a post-process of {\it ab~initio} calculations. The fully
self-consistent {\it ab~initio} calculations are performed for the
bulk compounds in the framework of density functional theory
(DFT)~\cite{DFT} using BSTATE (Beijing Simulation Tool for Atom
Technology) package~\cite{Fang} with the plane-wave ultra-soft
pseudo-potential method~\cite{Vanderbilt}. Generalized gradient
approximation (GGA) of the Perdew-Burke-Ernzerhof (PBE) type is used
for the exchange-correlation potential~\cite{Perdew}. To guarantee
the convergence, we use 340 eV for the cut-off energy of
wave-function expansion, and 10$\times$10$\times$10 K-points mesh in
the BZ. All structure parameters of Bi$_2$Se$_3$, Sb$_2$Te$_3$,
Bi$_2$Te$_3$ are chosen from experimental data listed in
Table~\ref{parameters}.

After completing the self-consistent {\it ab~initio} calculations,
two steps are followed to construct the PAWF basis and the effective
Hamiltonian. First we disentangle an isolated group of bands through
minimizing the invariant part of the spread functional of wave
functions ($\Omega_I$)~\cite{Souza}. In the second step, the Hilbert
space of this isolated group of bands is directly projected onto
atomic $p$ orbitals (keeping the angular symmetry). Finally a set of
unitary rotation matrices ${\cal U}({\bf k})$ can be obtained, which
can be straightly used to construct the PAWF and the effective
Hamiltonian.  The details are elucidated in the following.

\subsection {Extracting of isolated group of bands}

For a simple band insulator, the disentangling procedure is typically
not necessary, and it is straightforward to choose a set of Wannier
basis expanding the identical space of occupied Bloch states. However,
in our case, aiming to describe the electronic characters near the
Fermi level, both occupied and unoccupied bands are necessary, which
are not detached from higher bands.  By partial density of states
(PDOS) analysis, we know that the states near the Fermi level mainly
come from the contribution of $p$ orbitals of both Bi and Se
atoms. Since each atom has three $p$ orbitals, the total number of
orbitals of interest is $N = 15$, consisting of 9 valence bands and 6
unoccupied bands. Following the procedure described in
MLWF~\cite{Souza}, we introduce a big enough outer energy window (-20
eV, 20 eV), and a small inner windows (-6 eV, 2 eV). A smooth subspace
(of N bands) can then be constructed by optimizing the spread
functional, namely $\Omega_I$, as suggested in Ref.[\cite{Souza}].

\subsection {PAWF basis and effective Hamiltonian}

The Bloch functions $\psi_{m\bf{k}}$ of the isolated group of $N$
bands are directly projected onto the atomic $p$ orbitals, and the
unitary rotation matrices ${\cal U}({\bf k})$ can be obtained by
proper orthonormalization~\cite{Marzari,PWANF}. With these unitary
rotation matrices $ {\cal U}({\bf k})$, we can obtain well-localized
PAWF by rotating the original Bloch functions in the following way:
\begin{equation}
|\tilde{\psi}_{n\bf{k}}> = \sum_{m=1}^{N}{\cal U}_{mn}({\bf k})|\psi_{m\bf{k}}>,
\end{equation}
\begin{equation}
|W_n^R>=\frac{1}{N_k}\sum_ke^{-ikR}|\tilde{\psi}_{n\bf{k}}>,
\end{equation}
where $n$ and $m$=1,2 $\cdots$ $N$ are band index, $N_k$ is the total
number of $k$-points, $W_n^R$ denotes the $n$-th PAWF orbital, which
is centered at the lattice vector $R$.

Using the PAWF as basis, the effective Hamiltonian $H^W(k)$ can be
obtained correspondingly by rotating the original Hamiltonian,
\begin{equation}
H^{W}({\bf k}) = {\cal U}^{\dag}H({\bf k}){\cal U}.
\end{equation}
Because our PAWF are constructed from atomic orbitals, which have
atomic symmetry, we can further symmetrize the Hamiltonian using
their atomic characters. Using Fourier transformation, the PAWF
effective Hamiltonian in real space can be given by
\begin{equation}
H^{W}({\bf R}) = \frac{1}{N_{k}}\sum_{\bf k}e^{-i{\bf k\cdot R}}
H^{W}({\bf k}).
\end{equation}
The Hamiltonians at any other $k$-point can be now obtained by
transforming $H^{W}(\bf{R})$ back into $k$ space,
\begin{equation}\label{eq:tb}
H^W({\bf k}) = \sum_{\bf R} e^{i{\bf k\cdot R}} H^{W}({\bf R}).
\end{equation}

All the hopping matrix elements in the effective Hamiltonian are
directly calculated via constructing the PAWF. This approach differs a
lot from the conventional Slater-Koster TB
method~\cite{Slater_PR94_1954} where hopping integrals up to certain
neighbors are obtained by a fitting method.  The present method has
several special advantages. First, it is not necessary to maximize the
localization of Wannier functions (WF). Such step will usually break
the local symmetry of WF. In our case, the WF could be constrained to
satisfy local crystal symmetry, and the PAWF effective Hamiltonian can
be symmetrized, while keeping the necessary accuracy. Second, the
atomic spin-orbit coupling (SOC) can be easily implemented. Since the
WF are constructed from atomic orbitals, they are already quite
localized although not maximally localized by definition. The PAWF
effective Hamiltonian is expected to be useful for large supercell
calculations without extra errors from the symmetry problem.

\subsection{The spin-orbit coupling}

The SOC is important for the topological nature of Bi$_2$Se$_3$
family compounds, and should be included in all analysis. Since SOC
is mostly local atomic physics and has little \textbf{k}-dependence,
the simplest way to supplement SOC on top of our PAWF effective
Hamiltonian (denoted as $H^{0}$ from now on) is to include an
additional local term $H^{soc}$ in the total
Hamiltonian~\cite{Chadi},
\begin{equation} \label{eq:htot}
H^{tot} = H^{0} + H^{soc},
\end{equation}
where
\begin{equation}
H^{soc} = (\hbar/4m^{2}c^{2})[\nabla \bf{V} \times \bf{P}]\cdot
{\bf{\sigma}}.
\end{equation}
$H^{soc}$ comes from the SOC interaction, while {\bf V} and
${\bf\sigma}$ represent total potential and Pauli spin matrices
respectively.

As discussed above, our PAWF are constructed from atomic $p$ orbitals
(having atomic angular symmetry), the SOC Hamiltonian can be
straightforwardly written down, in terms of those PAWF basis,
\begin{eqnarray}
  |p_x,\uparrow>, |p_y,\uparrow>, |p_z,\uparrow>,|p_x,\downarrow>,
|p_y,\downarrow>, |p_z,\downarrow>,
\end{eqnarray}
where $\uparrow(\downarrow)$ indicates the spin. With this basis set
for each atom, $H^{soc}$ part can be expressed as
\begin{equation}
H^{soc}=\frac{\lambda}{2}\left[\begin{array}{cccccc}
0 & -i & 0 & 0 & 0 & 1 \\
i & 0 & 0  & 0 & 0 & -i \\
0 & 0 & 0  & -1 & i & 0 \\
0 & 0 & -1  & 0 & i & 0 \\
0 & 0 & -i  & -i & 0 & 0 \\
1 & i & 0   & 0   &0  &0 \end{array}\right].
\end{equation}
where $\lambda$ denotes the SOC parameter. We take the SOC
parameters of Bi, Se, Sb and Te atoms from Wittel's spectral data
($\lambda_{Bi}$ = 1.25 eV, $\lambda_{Se}$ = 0.22 eV, $\lambda_{Sb}$
= 0.4 eV, $\lambda_{Te}$ = 0.49 eV)~\cite{Wittel}.

\subsection{Accuracy of effective Hamiltonian: results for bulk}

\begin{figure*}
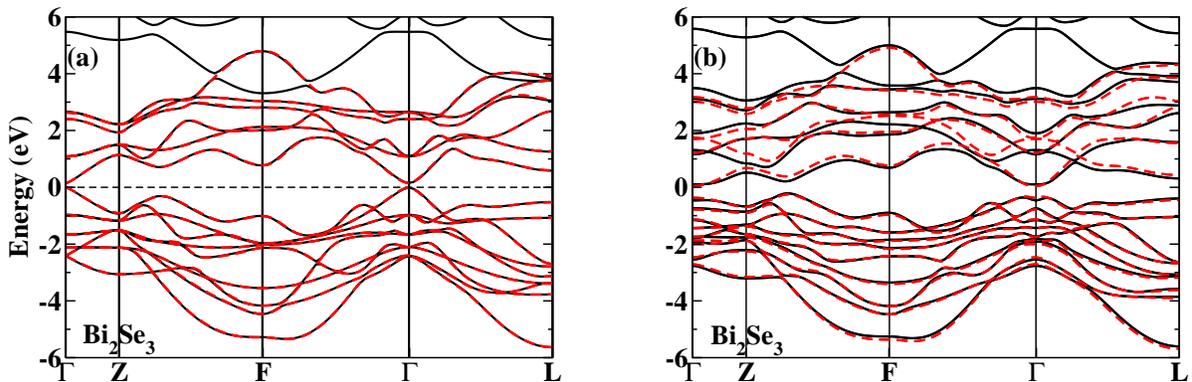

\includegraphics[clip,scale=0.3]{noSOCband.eps}
\ \ \ \ \ \ \ \ \ \ \ \ \ \ \ \
\includegraphics[clip,scale=0.3]{BSband.eps}
\caption{(Color online) The comparison of PAWF bands (red dashed
  lines) with the {\it ab~initio} band structures (black solid lines)
  of Bi$_2$Se$_3$. (a) The results without SOC. The PAWF effective
  Hamiltonian almost exactly reproduces the {\it ab~initio} band
  structures.  (b)The results with SOC for
  Bi$_2$Se$_3$.}\label{bandBS}
\end{figure*}

\begin{figure*}
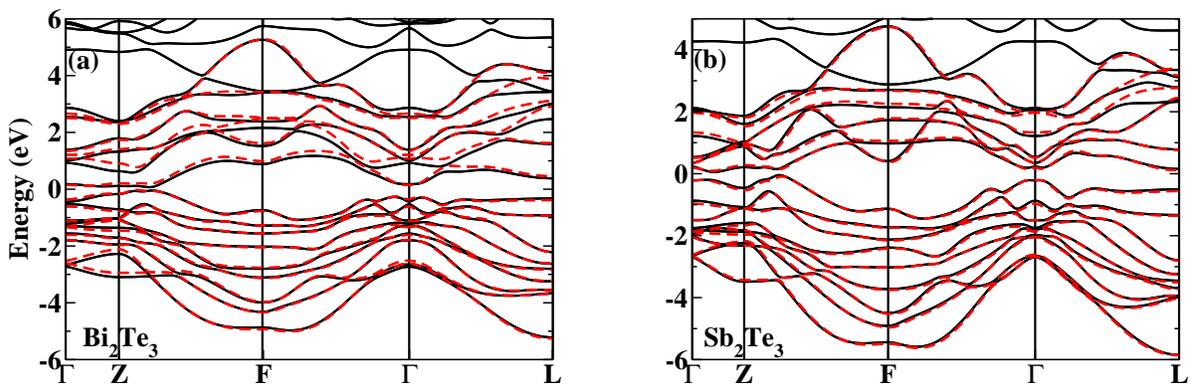

\includegraphics[clip,scale=0.3]{BTband.eps}
\ \ \ \ \ \ \ \ \ \ \ \ \ \ \ \
\includegraphics[clip,scale=0.3]{SBband.eps}
\caption{(Color online) The comparison of PAWF bands (red dashed
  lines) with the {\it ab~initio} band structures (black solid lines)
  of (a) Bi$_2$Te$_3$ and (b) Sb$_2$Te$_3$. Only the results after
  including the SOC interaction are shown here.}\label{bandBT}
\end{figure*}

To demonstrate the quality of the PAWF effective Hamiltonian, here
we compare several important properties calculated from the PAWF
Hamiltonian with that obtained from the $ab~initio$ calculations.
The first property to be compared is the the calculated bulk band
structure.  As shown in Fig.~\ref{bandBS}(a), for Bi$_2$Se$_3$
without SOC, the band structures obtained from PAWF effective
Hamiltonian can almost exactly reproduce the {\it ab~initio} band
structures. After taking into account the SOC interaction (as shown
in Fig.~\ref{bandBS}(b)), the quality is slightly reduced, but it is
still reasonably good since only the \textbf{k}-independent SOC
interaction is implemented in the effective Hamiltonian. For
Bi$_2$Te$_3$ and Sb$_2$Te$_3$, the PAWF effective Hamiltonian show
similar quality (shown in Fig.~\ref{bandBT}(a)-(b)).

\begin{table}
\caption{The parity eigenvalues of the energy bands of Bi$_2$Se$_3$
family compounds at $\Gamma$ point, obtained from the effective
  Hamiltonian. The parity numbers of nine occupied bands and the first
  conduction band are shown (the corresponding band energy increases
  from left to right). The parity products of nine occupied
  bands are given in brackets.}
\begin{tabular}{l c|c c c c c c c c c c}
\hline \hline
\multicolumn{12}{c}{Parity}\\
\hline
\hline
Bi$_2$Se$_3$   &(-1) &-1 &1  &-1  &-1 &1  &1  &-1 &-1 &1; &-1    \\
Bi$_2$Te$_3$   &(-1) &-1 &1  &1   &-1 &-1 &1  &-1 &-1 &1; &-1    \\
Sb$_2$Te$_3$   &(-1) &1  &-1 &-1  &1  &-1 &1  &-1 &-1 &1; &-1 \\
\hline
\end{tabular}\label{parity}
\end{table}

Parity is another important quantity to distinguish the topological
characters of Bi$_2$Se$_3$ family compounds. Because the present
systems posses the inversion symmetry, the method proposed by Fu and
Kane~\cite{Fu2} is used here. For all three systems, we calculate the
parity eigenvalues of nine occupied bands and the first conduction
band at the time-reversal invariant momentum (TRIM) points $\Gamma$,
$L$, $F$, and $Z$.  The results for the $\Gamma$ point, obtained from
the effective Hamiltonian, are listed in Table~\ref{parity}. As can be
clearly seen, the parity products of occupied bands are -1 for the
$\Gamma$ points. For the other TRIM points, $L$, $F$, and $Z$, the
detailed numbers are not listed here, but our results give +1 for
three of the compounds. From the parity numbers, we can therefore
identify the corresponding Z$_2$
invariants~\cite{Kane_PRL95_2005_146802,Fu1,Fu2} as 1 for all the
three compounds. This means they have nontrivial topological nature,
in good agreement with our previous studies~\cite{Zhang}. From above
comparisons, we conclude that the quality of the PAWF effective
Hamiltonian is as good as that obtained from the {\it ab~initio}
calculations.

\section{The properties of surface states}

\begin{figure*} [tbp]
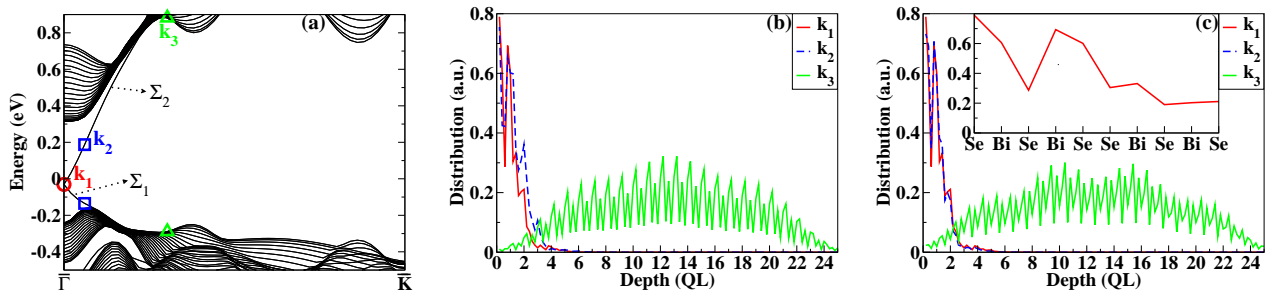

\includegraphics[clip,scale=0.21]{GKband.eps}
\ \ \ \ \
\includegraphics[clip,scale=0.21]{GKdis1.eps}
\ \ \ \ \
\includegraphics[clip,scale=0.21]{GKdis2.eps}
\caption{(Color online) (a) The calculated band structures of
  Bi$_2$Se$_3$ slab with film thickness of 25-QLs. The bands along the
  $\overline{\Gamma}\rightarrow \overline{K}$ direction are shown. The
  Fermi level is located at energy zero. The two surface bands in the
  bulk gap (around the $\overline{\Gamma}$ point) are denoted as
  $\Sigma_1$ and $\Sigma_2$. (b) and (c) The real-space distribution
  of eigen wave-functions ($|\psi_{nk}(\textbf{r})|^2$) for different
  $k$-points ($k_1$, $k_2$ and $k_3$ as indicated in (a)). (b) or (c)
  are for $k$-points along the $\Sigma_1$ or $\Sigma_2$ lines
  respectively. The inset of (c) is the zoomed-in picture for the
  $k_1$ point eigen wave function, where the atomic layer resolution
  is visible.}\label{GK}
\end{figure*}

\begin{figure*} [tbp]
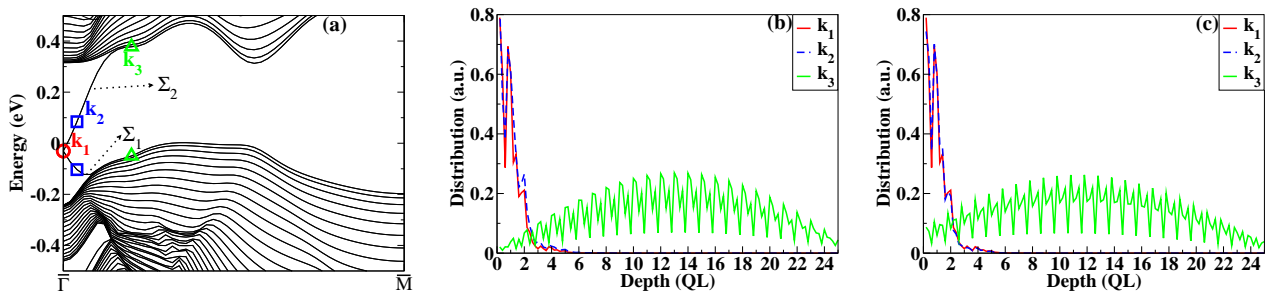

\includegraphics[clip,scale=0.21]{GMband.eps}
\ \ \ \ \
\includegraphics[clip,scale=0.21]{GMdis1.eps}
\ \ \ \ \
\includegraphics[clip,scale=0.21]{GMdis2.eps}
\caption{(Color online) The same as Fig.~\ref{GK}, but for bands along
  the $\overline{\Gamma}\rightarrow \overline{M}$
  direction.}\label{GM}
\end{figure*}

Once the effective Hamiltonian has been constructed from above
procedures, we will be able to calculate the surface states of
semi-infinite systems from the iterative Green's function
method~\cite{Sancho}, as described in our previous
studies~\cite{Zhang}. The existence of gapless spin-filter surface
states is the direct manifestation of the topological nature. The
crystal structure of Bi$_2$Se$_3$ family compounds can be understood
as the stacking of QLs along the $z$ direction. The inter-layer
bonding between two QLs is much weaker than that inside the QL, it is
natural to expect that the cleavage plane should be between two QLs.
This fact has been well confirmed by the recent experiments on the
layer-by-layer MBE growth of ultra-thin
film~\cite{Kehui,Xue1,Xue2,Xue3}.  We therefore focus our studies on
this type of surface termination (with Se1 atomic layer as the top
most layer). Our previous studies have shown that gapless surface
states with single Dirac point (located at $\overline{\Gamma}$) exist
for all three compounds, Bi$_2$Se$_3$, Bi$_2$Te$_3$ and Sb$_2$Te$_3$,
but not for Sb$_2$Se$_3$. The dispersion of surface states around
Dirac point is highly linear, and these surface states can be
described by a simple continuous model~\cite{Zhang}. Here, we will not
repeat those results, on the other hand however, will focus on some
other details of surface states, including the penetration depth, the
spin-resolved Fermi surface, and the chiral spin texture of the
surface states.

\subsection{The penetration depth of surface states}

The spread of surface states, or in other words its spacial
penetration depth into the bulk, is an essential quantity for the
potential applications of these surface states if any. In order to
clarify this point, a free standing slab model is constructed using
our PAWF effective Hamiltonian. We use a slab consisting of 25 QLs,
which is thick enough to avoid the direct coupling between the two
surfaces, i.e, the top and the bottom surfaces of the slab. In order
to make the calculations more realistic, we further take into account
the surface correction.  The bulk PAWF effective Hamiltonian is used
to construct the Hamiltonian of the slab, however for those layers
close to the surface, the on-site energies of PAWF should be modified
from its bulk counterparts due to the presence of vacuum (i.e, the
surface potential). By comparing with the fully self-consistent {\it
  ab~initio} calculations of thin slab, all the correction of on-site
energies for PAWF ($\Delta E_n = E_n^{surface} - E_n^{bulk}$) can be
obtained.

The calculated band structures of our 25-QLs Bi$_2$Se$_3$ slab along
$\overline{\Gamma}\rightarrow \overline{K}$ and
$\overline{\Gamma}\rightarrow \overline{M}$ lines are shown in
Fig.~\ref{GK} and Fig.~\ref{GM}. Clearly, within the bulk energy
gap, there exist two surface bands (marked by $\Sigma_1$ and
$\Sigma_2$), which are degenerate at $\overline{\Gamma}$ point
(where so called Dirac cone exists). The two surface bands
($\Sigma_1$ and $\Sigma_2$) are sampled by several $k$-points (with
$k_i$, i=1,2,3). The real space distribution of eigen wave functions
for those sampling $k$-points are plotted in Fig.~\ref{GK}(b)-(c)
and Fig.~\ref{GM}(b)-(c) respectively. For those $k$-points close to
the Dirac point (such as $k_1$ and $k_2$), the distribution of wave
functions are very localized to the surface region, with a typical
spread of about 2 QLs (about 2nm in thickness). By moving away from
the Dirac point, the penetration depth increases, and finally for
$k_3$ point, where the surface state almost merges with the bulk
states, the eigen wave function becomes a extended state. For both
$\overline{\Gamma}-\overline{M}$ and
$\overline{\Gamma}-\overline{K}$ lines, the same behaviors are
observed. To further see the oscillating behavior of surface state
distribution over atomic layers, we plot the distribution for $k_1$
in the inset.  Finally, we can conclude that surface states are very
localized to the surface region, and the penetration depth is about
2 or 3 QLs.

\begin{figure}[tbp]
\includegraphics[clip,scale=0.4]{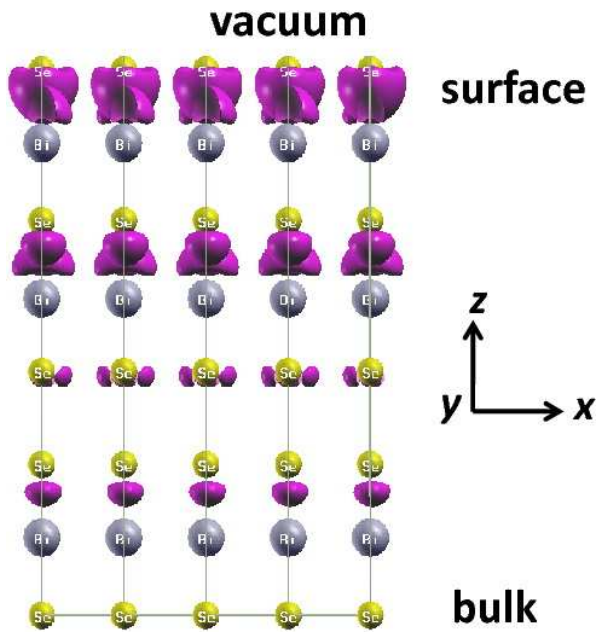}
\caption{(Color online) The spatial charge distribution within an
  energy window of 10meV around the Fermi level. The top atomic layer
  means the surface of the slab. It is clearly seen that electrons are
  mainly localized at the surface of the slab.}\label{figcharge}
\end{figure}

To make further comparison, we have performed a fully
self-consistent {\it ab~initio} calculations for a 3-QLs slab. The
real space charge distribution of surface states can thus be
obtained, by integrating the states within an energy window of 10
meV around the Dirac point. As shown in Fig.~\ref{figcharge}, the
charge distribution is really localized at the surface region.
Further more the charges are not centered at the atom, but rather
distributed mostly between the Bi and Se atoms, where strong
covalent bonding is expected.

\subsection{The spin-resolved Fermi surface}

Probing the $\pi$ Berry phase enclosed by the Fermi surface of surface
states is one of the most direct methods to distinguish TI. Now we
will analyze the spin-resolved Fermi surfaces around the Dirac point
of the semi-infinite Bi$_2$Se$_3$ system. With the Green's functions
calculated from the effective Hamiltonian, the spin-filter surface
states and the corresponding Fermi surfaces can be obtained directly.

\begin{figure}[tbp]
\includegraphics[clip,scale=0.2]{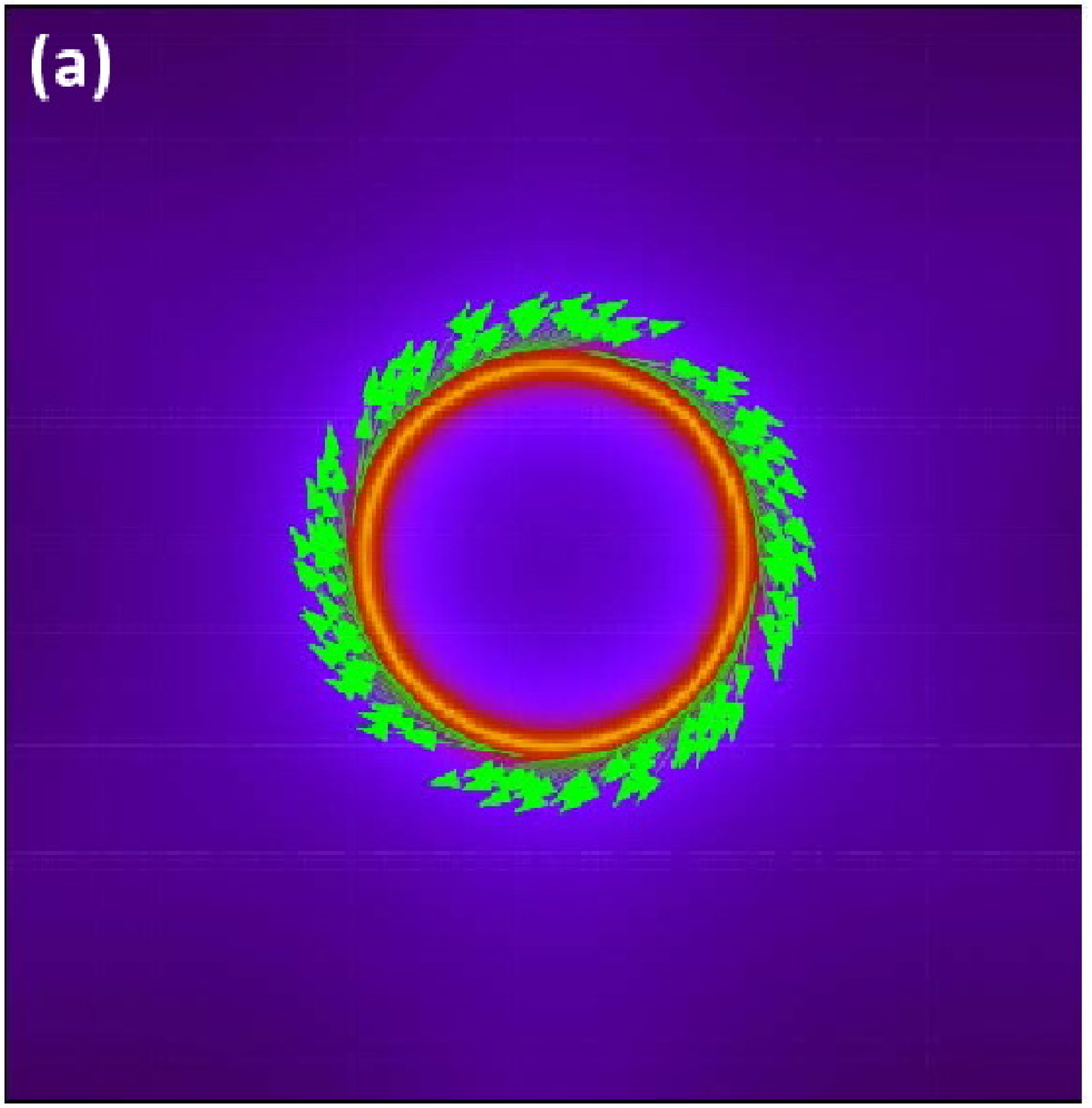}
\includegraphics[clip,scale=0.2]{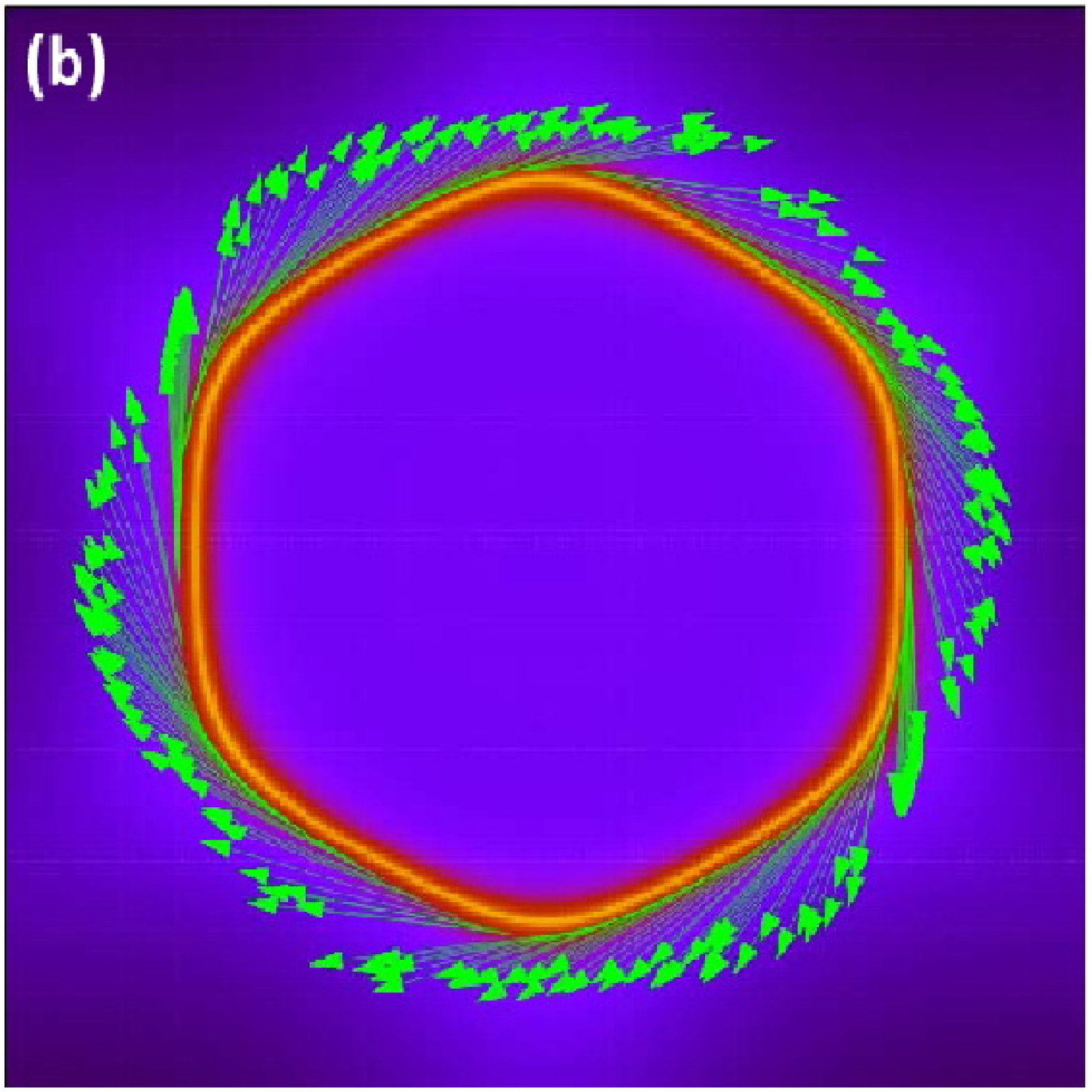}
\caption{(Color online) The spin resolved Fermi surface of surface
  states. (a) and (b) show the Fermi surfaces for Fermi energy at
  0.1eV and 0.25eV, respectively. The Fermi surfaces are denoted by
  red circle, and the in-plane spin orientation is indicated by green
  arrows.}\label{figFS}
\end{figure}

As shown in Fig.~\ref{figFS}, when the Fermi level is close to the
Dirac point, the corresponding Fermi surface is nearly a perfect
circle; while if the Fermi level is away from the Dirac point, the
properties of the surface states are significantly affected by the
bulk states and thus satisfy the crystal symmetry. For example, the
Fermi surface for $E_f$=0.1eV (shown in Fig.~\ref{figFS}(a)) looks
like a small circle, while the Fermi surface for $E_f$=0.25eV (shown
in Fig.~\ref{figFS}(b)) looks like a hexagon satisfying $c_{3z}$
symmetry. The spin orientation for states around the Fermi surface is
marked as well by arrows. The magnitude of the spin $z$ component is
very small, only about $1/40$ of the in-plane component. Thus the spin
almost completely lies in the plane. Moving around the Fermi surface,
the spin orientation rotate simultaneously, forming a spin-orbit ring,
which carries a $\pi$ Berry phase.  This signifies the topological
nontrivial properties of Bi$_2$Se$_3$. In addition, as shown in
Fig.~\ref{figFS}, the spin orientation of the ring belongs to the left
chirality (the normal direction of the semi-infinite system is defined
as z direction) which is again one of the important manifestations of
nontrivial topological characters.

\subsection{The linear dispersion of surface states}

\begin{figure}[btp]
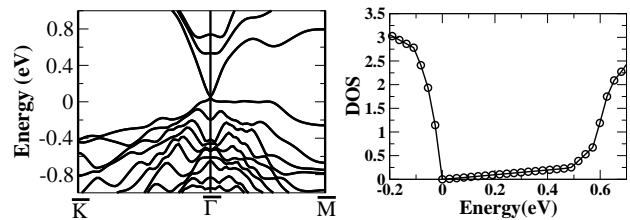

\includegraphics[clip,scale=0.33]{band_up.eps}
\includegraphics[clip,scale=0.3]{SS_dos.eps}
\caption{(a) Energy bands of a 3-QLs slab. (b) The corresponding
  density of states (DOS). The DOS near the Fermi level is highly
  linear to energy due to the presence of Dirac cone type surface
  states.}\label{figlinear}
\end{figure}

The gapless surface states that connect the bulk valence and
conduction bands have almost linear energy dependence near the Dirac
point. Linear band structure in 2D should lead to linear density of
states (DOS). To be specific, in the bulk energy gap, the energy bands
are mostly from the surface states, so the DOS in the bulk energy gap
are expected to be highly linear. Fig.~\ref{figlinear}(a) shows the
band structures of a 3-QLs slab. Its corresponding DOS (presented in
Fig.~\ref{figlinear}(b)) shows very nice linear energy dependence
within the bulk energy gap as expected. This type of DOS can be easily
measured by low temperature scanning tunneling spectroscopy (STS),
which will provide an indirect method to probe the existence of linear
surface states.

\section{3D topological phase transition}

The above discussions presented in sections II and III concentrate on
the effective Hamiltonian and the detailed understanding to the
properties of surface states. By such a way, we can analyze the
topological nature of specific compound. On the other hand, however,
the topological nature can be also understood from simple bulk
studies, where a gap close-reopening transition (a topological phase
transition) can be obtained by tuning some
parameters~\cite{Murakami}. We will show in this section that such a
phase transition can be indeed obtained in
Sb$_2$(Te$_{1-x}$Se$_x$)$_3$ alloy system.

\begin{figure}[tbp]
\includegraphics[clip,scale=0.34]{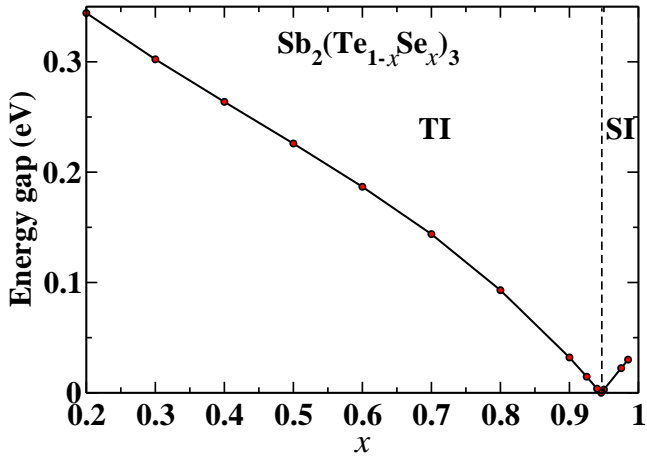}
\caption{(Color online)The Gap energy of Sb$_2$(Te$_{1-x}$Se$_x$)$_3$
  as a function of Se concentration $x$.  TI denotes the topological
  insulator, while SI means the simple insulator.}\label{figE3}
\end{figure}

The virtual crystal approximation (VCA) method proposed by
Bellaiche~\cite{Bellaiche} is used here to simulate the doping
process. The solid solution elements are treated as different atomic
species with their own weights, rather than being treated as a whole
virtual atom.  With this method, we investigate the Se doping
dependence of band structure of Sb$_2$(Te$_{1-x}$Se$_x$)$_3$. We
assume that crystal parameters and positions of inner atoms change
linearly with doping, and they are obtained by linear interpolation in
calculation. The evolution of the gap energy of
Sb$_2$(Te$_{1-x}$Se$_x$)$_3$ as a function of Se concentration is
calculated and illustrated in Fig.~\ref{figE3}. With increasing Se
doping, the valence band maximum and conduction band minimum get
closer gradually (0 $< x \leq$ 0.94), attributed to the gradually
decreased SOC strength. Consequently, the two bands cross at the
critical point $x$ = 0.94, resulting in a 3D topological quantum phase
transition. An band-inverse appears when the doping concentration
further increases. Therefore we can conclude that the occurrence of 3D
topological phase transition is driven by the SOC interaction. In
reality, the most stable crystal structure of Sb$_2$Se$_3$ is slightly
distorted from the rhombohedral structure of Sb$_2$Te$_3$, however, as
long as the Sb$_2$Se$_3$ is topologically trivial, such 3D topological
phase transition will be expected. At the phase transition point
(x=0.94 as predicted in our calculations), the 3D Dirac cone may be
expected in the bulk band structure~\cite{Murakami}.

\section{conclusions}

Bi$_2$Se$_3$, Bi$_2$Te$_3$, and Sb$_2$Te$_3$ systems are a new class
of TIs. In the present paper, we construct the effective Hamiltonian
for this family compounds based on the PAWF method. The effective
Hamiltonian can well reproduce the {\it ab~initio} band structures and
its topological nature. The penetration depth and the spin-resolve
Fermi surfaces of surface states are calculated and analyzed in
detail. At the end of the paper, we discuss the 3D topological phase
transition in Sb$_2$(Te$_{1-x}$Se$_x$)$_3$ and suggest that 3D Dirac
cone can be obtained by doping Se in Sb$_2$Te$_3$. Finally, we hope
that our effective Hamiltonian can provide further understanding of
these materials in future.

\section{ACKNOWLEDGMENTS}

We acknowledge the valuable discussions with C. X. Liu, X. L. Qi,
S. C. Zhang, S. Q. Shen, Q. Niu, and the supports from the NSF of
China, the 973 Program of China, and the International Science and
Technology Cooperation Program of China.

\end{document}